\def\expec#1{\langle#1\rangle}

\def\etal{{\frenchspacing\it et al.}}
\def\ie{{\frenchspacing\it i.e.}}
\def\eg{{\frenchspacing\it e.g.}}
\def\etc{{\frenchspacing\it etc.}}
\def\rms{{\frenchspacing r.m.s.}}

\def\pp{\noindent\parshape 2 0truecm 13.6truecm 1truecm 12.6truecm}
\def\rf#1;#2;#3;#4 {\par\pp#1, {\it #2}, {\bf #3}, #4. \par}
\def\rg#1;#2;#3;#4;#5 {\par\pp#1, {\it #2}, {\bf #3}, #4 (``#5"). 
\par}
\def\rn{\pp}

\def\beq#1{\begin{equation}\label{#1}}
\def\eeq{\end{equation}}
\def\beqa#1{\begin{eqnarray}\label{#1}}
\def\eeqa{\end{eqnarray}}
\def\eq#1{equation~(\ref{#1})}
\def\Eq#1{Equation~(\ref{#1})}
\def\eqnum#1{~(\ref{#1})}

\def\bfig{\begin{figure}[h] \centerline{\hbox{}}\vfill}
\def\efig{\end{figure}\vfill\newpage}

\def\fheight{12cm}
\def\fwidth{17cm}
\def\fig#1{Figure~\ref{#1}}
\def\Fig#1{Figure~\ref{#1}}

\def\spose#1{\hbox to 0pt{#1\hss}}
\def\simlt{\mathrel{\spose{\lower 3pt\hbox{$\mathchar"218$}}
     \raise 2.0pt\hbox{$\mathchar"13C$}}}
\def\simgt{\mathrel{\spose{\lower 3pt\hbox{$\mathchar"218$}}
     \raise 2.0pt\hbox{$\mathchar"13E$}}}
\def\simpropto{\mathrel{\spose{\lower 3pt\hbox{$\mathchar"218$}}
     \raise 2.0pt\hbox{$\propto$}}}

\def\addr#1{{\small\it #1}}

\def\expec#1{\langle#1\rangle}

\def\Mpc{{\rm Mpc}}
\def\muK{\mu\hbox{K}}
\def\nbar{{\bar n}}
\def\ntil{{\tilde n}}
\def\rh{\widehat{{\bf r}}}
\def\l{\ell}

\def\alm{a_{\l m}}
\def\alpmp{a_{\l'm'}}

\def\Dlm{\Delt_{\l m}}
\def\Dlpmp{\Delt_{\l'm'}}
\def\Cl{C_\l}
\def\Ylm{Y_{\l m}}
\def\Ylpmp{Y_{\l'm'}}

\def\flux{\phi}
\def\fluxc{{\flux_c}}
\def\m{m}
\def\Delt{\Delta}
\def\x{x}
\def\B{B}
\def\f{f}
\def\Cps{C^{ps}}
\def\Clmag{C_{\l}^{mag}}
\def\d{d}
\def\N{N}
\def\tento#1{\times 10^{#1}}

\def\ed{\end{document}}

\documentstyle[11pt,epsf,rotate]{article}
\begin{document}


\begin{titlepage}   

\noindent
March 4, 1997
\begin{center}

\vskip0.9truecm
{\bf

IS LENSING OF POINT SOURCES A PROBLEM FOR FUTURE
CMB EXPERIMENTS?\footnote{
Accepted for publication in {\it MNRAS}.
Available from\\
{\it h t t p://www.sns.ias.edu/$\tilde{~}$max/lensing.html} 
(faster from the US) and from\\
{\it h t t p://www.mpa-garching.mpg.de/$\tilde{~}$max/lensing.html} 
(faster from Europe).\\
Note that figures 1 and 3 will print in color if your printer supports it.
}
}

\vskip 0.5truecm

Max Tegmark$^{1,2,3,}$\footnote[4]{Hubble Fellow}
\&
Jens V. Villumsen$^3$

\smallskip
\addr{$^1$Institute for Advanced Study, 
Olden Lane, Princeton, NJ 08540; max@ias.edu}\\
\addr{$^2$Max-Planck-Institut f\"ur Physik, 
F\"ohringer Ring 6, D-80805 M\"unchen}\\
\addr{$^2$Max-Planck-Institut f\"ur Astrophysik, 
Karl-Schwarzschild-Str. 1, D-85740 Garching; jens@mpa-garching.mpg.de}\\

  \smallskip
  \vskip 0.2truecm

\end{center}

\begin{abstract} 
Weak gravitational lensing from large-scale structure 
enhances and reduces the fluxes from extragalactic
point sources with an {\rms} amplitude of order 15\%. 
In cosmic microwave background (CMB) experiments, 
sources exceeding some flux threshold $\fluxc$ are removed,
which means that lensing will modulate the brightness map 
of the remaining unresolved sources. Since this mean 
brightness is of order $100\muK$ at 30 GHz for a reasonable flux cut,
one might be concerned that this modulation 
could cause substantial problems for future CMB experiments. 
We present a detailed calculation of this effect 
and, fortunately, find that its power spectrum is 
always smaller than the normal point source power spectrum.
Thus although this effect 
should be taken into account when analysing future high-precision
CMB measurements, it will not substantially reduce the accuracy with
which cosmological parameters can be measured.
\end{abstract}

\end{titlepage}


\section{INTRODUCTION}

\begin{figure}[phbt]
\centerline{\hbox{$\>$}}\vskip-3truecm
\centerline{\epsfxsize=17cm\epsfbox{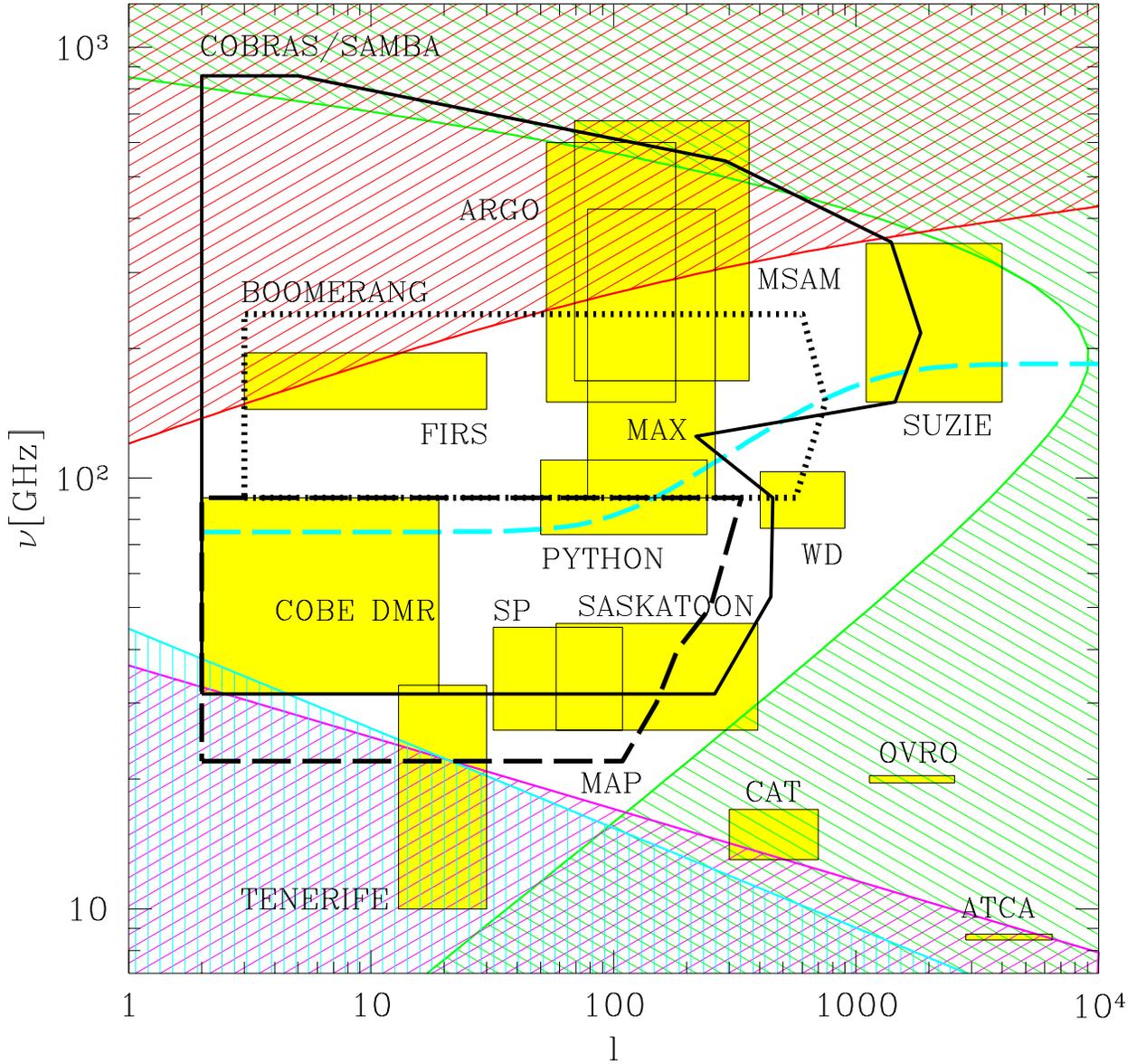}}
\caption{Where various foregrounds dominate.}
The shaded regions indicate where the various foregrounds cause 
fluctuations exceeding those of COBE-normalized scale-invariant
fluctuations, 
thus posing a substantial challenge
to estimation of genuine CMB fluctuations. 
They correspond to dust (top), free-free emission (lower left), 
synchrotron radiation (lower left, vertically shaded)
and point sources (lower and upper right).
The heavy dashed line shows the frequency where the total foreground
contribution to each multipole is minimal.
The boxes roughly indicate the range of multipoles $\l$ and frequencies
$\nu$ probed by various CMB experiments, as in TE96.
\label{EverythingFig}
\end{figure}

Future high-precision measurements of the fluctuations in the
cosmic microwave background (CMB) may allow key cosmological 
parameters (such as $\Omega$, $\Lambda$, $\Omega_b$, 
the Hubble constant, {\etc}) to be measured with 
unprecedented accuracy (Jungman {\etal} 1996;
Bond {\etal} 1997; Zaldarriaga {\etal} 1997),
and a number of ground-, balloon- and space-based missions are 
currently being planned for this purpose. 
Both when designing such missions and when analyzing the data
sets that they produce, it is important that all relevant
sources of foreground contamination are well understood,
to minimize the risk that foreground signals
are misinterpreted as CMB fluctuations. 
Since numerous experiments are are currently in the planning 
and design stages, it is therefore timely to catalog and quantify
all possible foreground effects.
\Fig{EverythingFig} summarizes recent 
foreground estimates from Tegmark \& Efstathiou (1996), 
hereafter ``TE96", and Bersanelli {\etal} (1996).
The purpose of this {\it Letter} is to investigate 
yet another foreground effect which has not been 
previously studied: that of weak gravitational lensing 
of point sources. 
In other words, we will discuss the extent to which 
the point source region (which occupies mainly the lower right
corner of \Fig{EverythingFig}) is expanded by lensing.

\begin{figure}[phbt]
\centerline{\rotate[r]{\vbox{\epsfysize=16cm\epsfbox{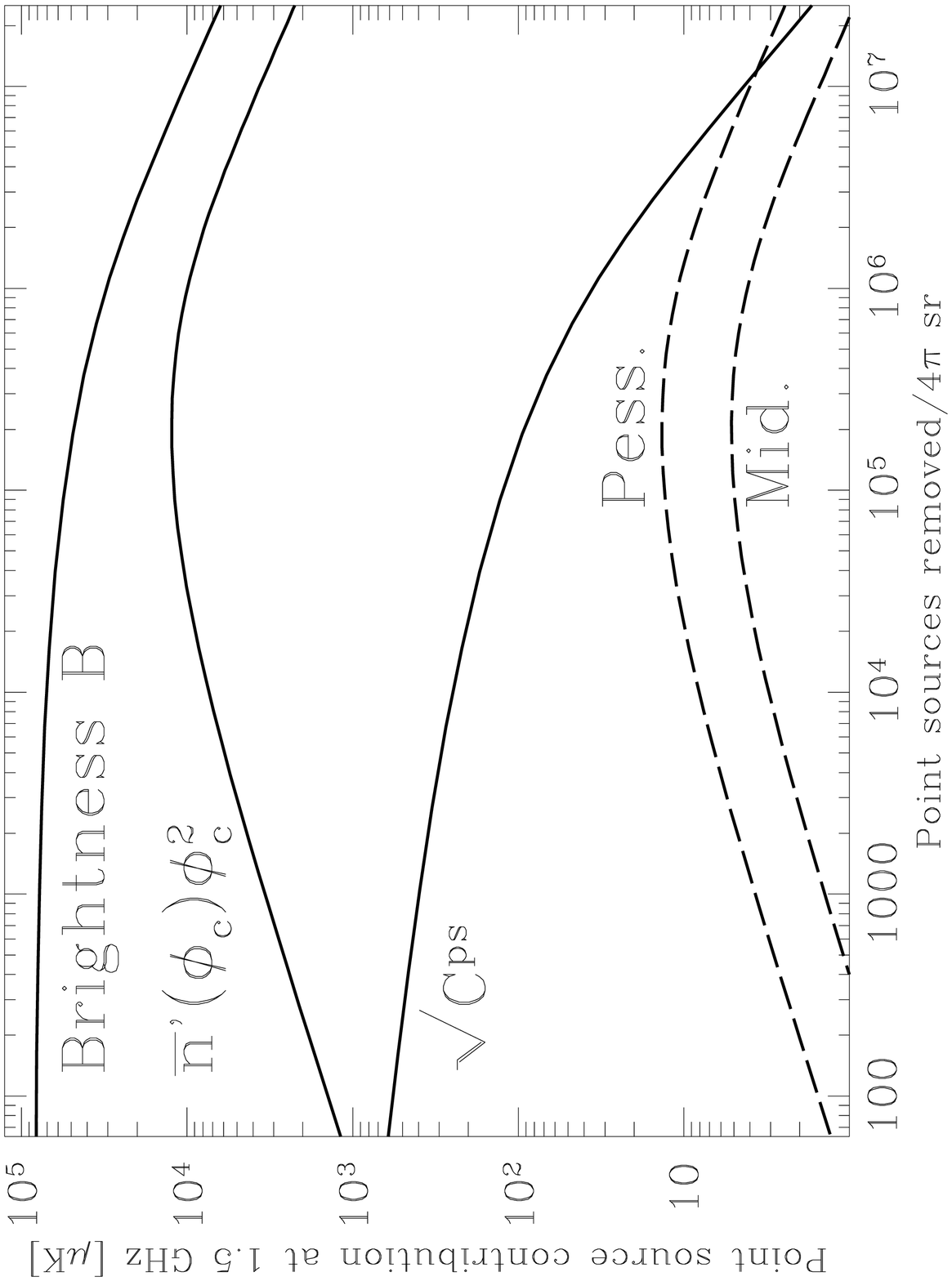}}}}
\caption{Dependence of radio source fluctuations on flux cut.}
\label{RemovalFig}
The three terms that contribute to
the point source power spectrum in \eq{PowerEq2} 
are plotted as a function
of the number of sources removed in an all-sky survey
(only the number per steradian matters),
based on the VLA FIRST source counts.
These terms (solid curves) are the integrated brightness of the population,
the edge effect at the flux cut and the shot noise 
fluctuations, respectively.
The dashed curves show the lensing contribution to 
$C_\l^{1/2}$ for the pessimistic and middle-of-the-road
scenarios, at the worst-case $\l$-values.
\end{figure}

Since it is impossible to remove all point sources
(as there are for all practical purposes infinitely many,
and not all with the same spectra), 
the standard procedure in CMB experiments is to remove
all point sources below some flux cut $\fluxc$, either by 
subtracting their estimated emission or by discarding all 
contaminated pixels. 
The total sky brightness from the remaining sources 
tends to be dominated by faint ones, 
whereas the brightness fluctuations (which 
contribute to \Fig{EverythingFig}) are dominated
by the sources just below the flux cut. For this reason,
the average brightness $\B$ exceeds 
the {\rms} fluctuations by a substantial factor, typically 
$\sim 10^3$, as shown in \fig{RemovalFig}. 
A process that caused even minor
spatial modulations of $\B$ could therefore pose a serious
problem for CMB experiments.

Weak gravitational lensing by large-scale structure 
(see Villumsen 1996 for a recent discussion)
can affect observed power spectra in more ways than one.
If we imagine a pattern painted on the inside 
of a rubber (celestial) sphere, weak lensing corresponds 
to stretching and compressing the rubber in
a random fashion, much like the way our mirror images get distorted
by non-flat mirrors in amusement parks.
For weak lensing, this distortion will
always constitute a one-to-one mapping of the image, {\ie}, 
there will be no caustics where the image ``folds over" on itself.
There is thus no smearing involved, so the fluctuation 
power is conserved. 
Instead, there is a ``Robin Hood effect" where power is redistributed
between multipoles, from those which have more to those which have
less (Seljak 1996).
For the CMB power spectrum, 
this effect is typically a few percent, and may well be detectable in 
future CMB experiments. 
For point sources, on the other hand, this effect is completely absent:
since they have a Poisson power spectrum with $\Cl$ constant, 
the Robin Hood effect (which effectively simply smoothes the power
spectrum), will leave $\Cl$ unaffected.

However, the presence of a flux cut for point sources produce a
different effect, which is absent for the CMB. 
In those regions where lensing causes magnification, we will see
(and remove) disproportionately many sources above the flux cut
(Turner {\etal} 1984, Broadhurst {\etal} 1995). 
Since lensing leaves the average brightness unaffected, 
the sky brightness from unremoved sources will be lower than average in 
this region. In other words, weak lensing causes the 
type of modulation of $\B$ that we warned about above.
The rest of this {\it Letter} is organized as follows. 
We present a detailed calculation of this effect in 
Section 2, assess its importance in Section 3 and summarize our 
findings in Section 4.

\section{CALCULATION OF THE EFFECT}

\subsection{How magnification bias affects the source counts}

Let the {\it unlensed source count function} $\nbar(>\flux)$ denote the
number of sources per steradian whose flux would exceed $\flux$
in the absence of lensing. 
When weak gravitational lensing magnifies a patch of sky with
a magnification factor $\m$, this has two separate effects on 
$\nbar(>\flux)$:
\begin{itemize}
\item The sources become a factor $\m$ brighter.
\item The sources are seen further apart, reducing their number density
by a factor $\m$.
\end{itemize}
In summary, the {\it lensed source count function}, which 
we denote $\ntil(>\flux)$, is given by
\beq{LensedSourceCountEq}
\ntil(>\flux) = {1\over\m}\nbar\left(>{\flux\over\m}\right).
\eeq
Defining the {\it differential source count} as 
$\nbar'(\flux)\equiv d\nbar(<\flux)/d\flux = -d\nbar(>\flux)/d\flux$
(a positive function), we thus have
\beq{LensedSourceCountEq2}
\ntil'(\flux) = {1\over\m^2}\nbar'\left({\flux\over\m}\right).
\eeq
For weak lensing, we can write $m=1+\Delt$, where $|\Delt|\ll 1$
(Broadhurst {\etal} 1995),
and we will make the approximation of dropping all quadratic 
and higher order terms in $\Delt$. 
Using $\nbar'(\flux/\m)\approx \nbar'(\flux-\flux\Delt)
\approx \nbar'(\flux) - \nbar''(\flux)\flux\Delt$, we can rewrite
\eq{LensedSourceCountEq2} as
\beq{LensedSourceCountEq3}
\ntil'(\flux,\rh) = \nbar'(\flux) - 
{1\over\flux}\left[\nbar'(\flux)\flux^2\right]'\Delt(\rh),
\eeq
where we have explicitly indicated the fact that the magnification
field depends on $\rh$, the unit vector pointing in our
direction of observation.

Since the total sky brightness 
$\int_0^\infty\nbar'\flux d\flux$ is finite, 
the quantity $\nbar'\flux^2$ must clearly approach zero both as 
$\flux\to 0$ and as $\flux\to\infty$, otherwise the integral would
diverge at the faint or bright end. Since the second term in
\eq{LensedSourceCountEq3} becomes a total differential when multiplied
by $\flux$, this means that the total brightness is not affected by 
lensing. This well-known fact is also verified by 
integrating $\phi$ times 
\eq{LensedSourceCountEq2} and changing variables.
However, when the upper integration limit is not $\infty$, 
the same procedure shows that the brightness contribution 
from all sources below some fixed flux cut {\it is} affected by lensing, 
and we will now evaluate this effect.

\subsection{The effect on the point source correlation function}

For CMB purposes, we can to a good approximation (TE96) 
model the locations of extragalactic point sources as completely 
uncorrelated. In the absence of lensing, we can
therefore write the observed density of point sources above some flux cut
$\fluxc$ as a sum of angular delta functions, 
$n(\rh)=\sum \delta(\rh,\rh_i)$, 
and model $n(\rh)$ as a Poisson process. 
A Poisson process satisfies
(see {\eg} Appendix A of Feldman, Kaiser \& Peacock 1994)
\beqa{PoissonEq1}
\expec{n(\rh)}_p&=&\nbar(\rh),\\
\label{PoissonEq2}
\expec{n(\rh)n(\rh')}_p&=&\nbar(\rh)\nbar(\rh')+
\delta(\rh,\rh')\nbar(\rh).
\eeqa
We use $\expec{\>}_p$ to denote ensemble averages with respect
to the Poisson process, to distinguish these from ensemble averages
with respect to the random field $\Delt$, which we will denote by
$\expec{\>}_f$. When performing both averages below, we will omit subscripts
and write $\expec{\>} = \expec{\expec{\>}_p}_f$.

Magnification bias forces us to take into account that
the total source population is the union of a 
number of independent subpopulations, 
corresponding to different flux classes.
We therefore write
\beq{SubpopEq}
n(\rh) = \int_0^\fluxc n'(\flux,\rh)d\flux,
\eeq
where $n'(\flux,\rh)d\flux$ is a sum of delta functions corresponding
to the sources whose fluxes fall between
$\flux$ and $\flux+d\flux$.
With this notation, equations\eqnum{PoissonEq1} and\eqnum{PoissonEq2} 
become generalized to 
\beqa{PoissonEq3}
\expec{n'(\flux,\rh)}_p&=&\ntil'(\flux,\rh),\\
\label{PoissonEq4}
\expec{n'(\flux,\rh)n'(\flux',\rh')}_p&=&\ntil'(\flux,\rh)\ntil'(\flux',\rh')+
\delta(\flux-\flux')\delta(\rh,\rh')\ntil'(\flux,\rh).
\eeqa
The observed sky brightness $\x$ in a direction $\rh$ is clearly 
given by
\beq{BrightnessFieldEq}
\x(\rh)\equiv\int_0^\fluxc n'(\flux,\rh)\flux d\flux,
\eeq
and we will now calculate its statistical properties.

Using 
equations\eqnum{LensedSourceCountEq3},\eqnum{PoissonEq3} 
and\eqnum{BrightnessFieldEq}, we see that the mean
is given by 
\beq{MeanEq}
\expec{\x(\rh)} 
= \int_0^\fluxc \expec{\expec{n'(\flux,\rh)}_p}_f \flux d\flux  
= \B - \expec{\f(\rh)}_f,
\eeq
where we have defined 
\beq{fDefEq}
\f(\rh) \equiv \nbar'(\fluxc)\fluxc^2 \Delt(\rh)
\eeq
and where $\B$ denotes the average total brightness due to sources below
our flux cut, {\ie},
\beq{BdefEq}
B\equiv \int_0^\fluxc\nbar'(\flux)\flux d\flux.
\eeq
Since $\expec{\Delt(\rh)}_f=0$ (there is just as much positive as negative
weak lensing from large-scale structure), the second term
in \eq{MeanEq} vanishes, and 
we see that regardless of the flux cut, lensing has no impact on the 
average brightness of unremoved point sources.

We now turn to the correlation function.
Using $\expec{\Delt(\rh)}_f=0$ and
equations\eqnum{LensedSourceCountEq3},\eqnum{PoissonEq3},\eqnum{PoissonEq4} 
and\eqnum{BrightnessFieldEq}, we obtain 
\beqa{CovCalcEq}
\nonumber
\expec{\x(\rh)\x(\rh')} 
&=& \int_0^\fluxc \int_0^\fluxc
\expec{\expec{n'(\flux,\rh) n'(\flux',\rh')}_p}_f 
\flux \flux' d\flux  d\flux'\\
&=& \expec{[\B-\f(\rh)][\B-\f(\rh')]}_f + \delta(\rh,\rh')\Cps,
\eeqa
where
\beq{CpsEq}
\Cps \equiv 
\int_0^\fluxc \nbar'(\flux)\flux^2 d\flux
\eeq
is the familiar power spectrum of unresolved point sources 
in the absence of lensing that was derived in TE96.
Using $\expec{f(\rh)}_f=0$ again, this reduces to 
\beq{CovEq}
\expec{\x(\rh)\x(\rh')} 
= \B^2 + \Cps + \nbar'(\fluxc)^2\fluxc^4\expec{\Delt(\rh)\Delt(\rh')}_f.
\eeq
Since the first term is merely the familiar and 
uninteresting monopole, the 
new effect that we have computed corresponds to the third term.
Thus the effect of lensing is to add
to the original point source correlation function
a new term which is the magnification correlation
times $\nbar'(\fluxc)^2\fluxc^4$.

\subsection{The effect on the point source power spectrum}

Let us expand the sky brightness $\x$ in spherical harmonics and
investigate the statistical properties of the expansion coefficients
\beq{almDefEq}
\alm\equiv \int\Ylm^*(\rh)\x(\rh) d\Omega.
\eeq
Assuming that the statistical properties of the 
magnification field $\Delt$ are isotropic (there is no need to
assume that $\Delt$ is Gaussian), its corresponding expansion coefficients
$\Dlm$ must satisfy
\beq{ClmagDefEq}
\expec{\Dlm^*\Dlpmp}_f = \delta_{\l\l'}\delta_{mm'}\Clmag
\eeq
for some $\Clmag$ which we will refer to as the 
{\it magnification power spectrum}.
Substituting \eq{CovEq} into 
\beq{PowerCalcEq}
\expec{\alm^*\alpmp} 
= \int\int \Ylm^*(\rh)\Ylpmp(\rh')\expec{\x(\rh)\x(\rh')} d\Omega d\Omega',
\eeq
we obtain
\beq{PowerEq1}
\expec{\alm^*\alpmp} = \delta_{\l\l'}\delta_{mm'}\Cl,
\eeq
where the point source power spectrum is
\beq{PowerEq2}
\Cl = 4\pi\delta_{0\l}\delta_{0m} B^2 
+ \Cps + \nbar'(\fluxc)^2\fluxc^4 \Clmag.
\eeq
The first two terms are merely the monopole 
and the shot noise power
(which is independent of $\l$), derived in TE96,
so the new effect due to lensing is given by the third term.

\section{HOW IMPORTANT IS THE NEW EFFECT?}

\subsection{Some useful approximations}

To clarify the relative importance of the last two terms
in \eq{PowerEq2}, it is convenient to express them in
terms of quantities that are more directly linked to 
observations. 
For typical scenarios, the differential source count
$\nbar'(\flux)$ is a smooth function on a log-log-plot, 
which means that near the flux cut $\fluxc$, we can approximate
it with a power law
\beq{PowerLawEq}
\nbar'(\flux) \approx \nbar'(\fluxc)\left({\flux\over\fluxc}\right)^{-\beta}
\eeq
for some constant $\beta$. 
To avoid the above-mentioned divergence of the total brightness,
we must have a logarithmic slope $\beta<2$ at the
faint end and $\beta>2$ at the bright end.
For instance, for radio sources at
1.5 GHz, the VLA FIRST survey 
gives the logarithmic slope 
$\beta\sim 1.6$ at $\fluxc\sim$ 1 mJy, steepening
to $\beta\sim 2.5$ at the bright end
(White {\etal} 1996, TE96) just as one would expect if 
the brightest sources are at low redshifts
$z\ll 1$ where evolutionary and cosmological 
effects are negligible, and the same qualitative behavior
is found at higher frequencies
(Windhorst {\etal} 1985, 1993).
Using \eq{PowerLawEq}, we thus find that
the expected number $\N$ of sources above the flux cut is
\beq{NdefEq} 
\N\equiv 4\pi\int_\fluxc^\infty \nbar'(\flux)d\flux \approx 
{4\pi\over\beta-1}\fluxc\nbar'(\fluxc)
\propto\fluxc^{-(\beta-1)},
\eeq
since the integral is dominated by sources just above the flux cut
for which the power law fit is very accurate.
Similarly, the integral in \eq{CpsEq} is dominated
by sources just below the cut, so using 
equations\eqnum{PowerLawEq} and\eqnum{NdefEq} 
gives 
\beq{CpsEq2}
\Cps \approx 
\left({\beta-1\over 3-\beta}\right)\left({N\over 4\pi}\right)\fluxc^2
\propto \fluxc^{3-\beta}.
\eeq
Finally, we can rewrite the lensing term in \eq{PowerEq2} as
\beqa{PowerEq3}
\nonumber
\nbar'(\fluxc)^2\fluxc^4 \Clmag
&\approx& 
(\beta-1)^2\left({N\over 4\pi}\right)^2\fluxc^2 \Clmag\\
\nonumber
&\approx& (\beta-1)(3-\beta)\left({N\over 4\pi}\right)\Clmag\Cps\\
&\propto&\fluxc^{-2(\beta-2)}.
\eeqa
\Eq{PowerEq2} showed that that power spectrum 
was a sum of three different terms. These are all plotted 
in \Fig{RemovalFig} as a function of the flux cut, 
using the source counts $\nbar'(\phi)$ from the VLA FIRST 
survey (White {\etal} 1996).
The approximations above allow us to understand 
all qualitative features of this figure.
Comparing equations\eqnum{CpsEq2} and\eqnum{PowerEq3}, we
notice that although the Poisson term
$\Cps$ {\it decreases} if we remove more sources, 
the lensing term {\it increases} as long as $\beta>2$, 
{\ie}, as long as we remove less than $\sim 10^5$ 
sources from an all-sky survey, or $\sim 4$ sources
per square degree.
This corresponds to $\sim 60$ sources in a 
$4^\circ\times 4^\circ$ field of the upcoming 
Very Small Array ({\it VSA}) experiment, and to a 
50 mJy flux cut at 1.5 GHz.

\subsection{The magnification power spectrum}

\begin{figure}[phbt]
\centerline{\epsfxsize=16cm\epsfysize=16cm\epsfbox{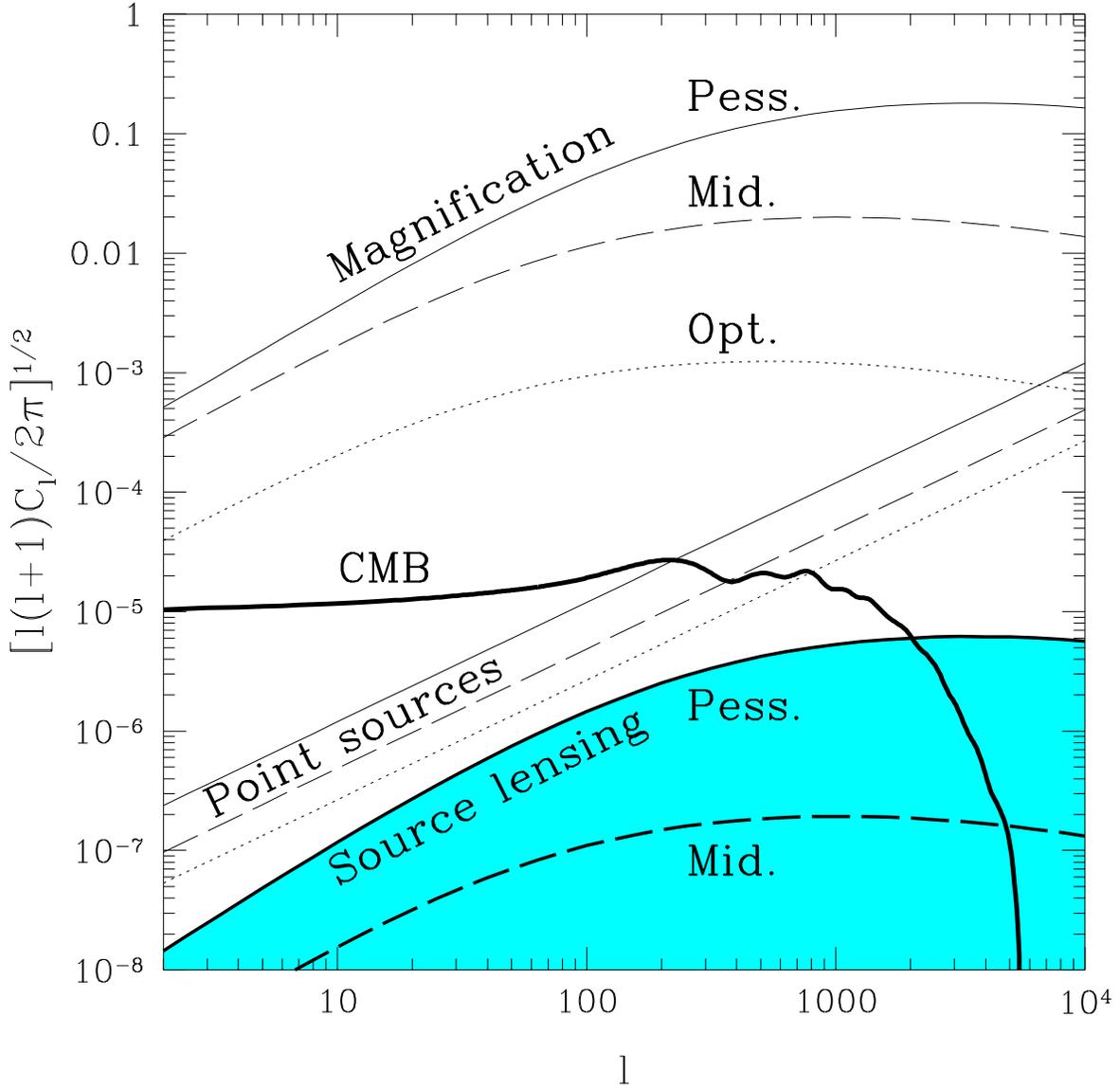}}
\caption{
The power spectra are plotted for the CMB (heavy line), 
the lensing magnification field (the three top lines), 
discrete point sources at 30 GHz (the three straight lines) and 
the lensing effect on them (the two bottom lines), all
for a standard CDM model. 
The solid, dashed and dotted lines correspond to the
pessimistic, middle-of-the-road and optimistic scenarios,
respectively.
}
\label{ClFig}
\end{figure}

\Eq{PowerEq3} shows that when $\beta\approx 2$, 
the ratio between the lensing term and the
conventional shot noise term $\Cps$ is simply 
$\Clmag$ times $N/4\pi$, the number of removed sources 
per steradian. 
$N/4\pi$ is typically $\gg 1$, and we will now evaluate 
$\Clmag$.
For any flat 
Friedman-Robertson-Walker cosmology, including non-linear clustering
and an arbitrary source distribution $n(x_s)$, the magnification power
spectrum is given by (Villumsen 1996)
\beq{JensEq1}
\Clmag =72\pi^3\Omega_0^2 d^{-3}\int_0^{x_h} w^2(x_l) (1+z)^2  P(\l/yd,z)\ dx_l,
\eeq
where the function $w$ is defined as
\beq{JensEq2}
w(x_l)\equiv \int_{x_l}^{x_h}  {y_{ls} \over y_s} n(x_s)\ dx_s.
\eeq
Here $\d\equiv H_0^{-1}c\approx 3000h^{-1}\Mpc$, 
$P(k,z)$ is the conventional three-dimensional matter power spectrum
at redshift $z$, and
$xd$ and $yd$ are the comoving radial and angular
distances to the epoch $z$, respectively. Subscripts $h$, $l$ and $s$ 
refer to horizon, lens and source positions, respectively, 
and $y_{ls}$ is the comoving angular lens-source distance.
These equations are valid on angular scales where the
sky is approximately flat, {\ie}, $\l\gg 60$, which is 
of course the
case in our regime of interest. 
For the simple case where $\Omega=1$, $\Lambda=0$, the density
fluctuations grow according to linear theory and all sources
are located at some characteristic redshift $z_s$,
this reduces to
\beq{JensEq3}
\Clmag =72\pi^3 d^{-3}
\int_0^{x_s} \left(1-{x\over x_s}\right)^2 P\left({\l\over xd}\right)\,dx,
\eeq
where $P$ is the current power spectrum and
$x_s = 2-2/\sqrt{1+z_s}$.
\begin{table}
\begin{center}
\begin{tabular}{|l||c|c|c|} 
\hline
Parameter			&Opt.	&Mid.	&Pess.\\
\hline
Source redshift			&0.4	&1	&$\gg$1\\
$\alpha$			&0.2	&0	&$-$0.3\\
$\Omega_0\sigma_8$		&0.2	&0.7	&1\\
Removed/sq deg			&0.5	&2	&4\\
$\beta(\fluxc)$				&2.5	&2.2	&2\\
\hline
\end{tabular}%
\caption{Parameters used in optimistic, middle-of-the-road and
pessimistic scenarios.}
\label{ScenarioTable}
\end{center}
\end{table}
This is plotted in \fig{ClFig} (dashed curves) 
for a standard CDM power spectrum
(Bond \& Efstathiou 1984) with $h=0.5$,
for the three cases described in 
Table 1. We label these scenarios 
``pessimistic", ``middle-of-the-road" and ``optimistic",
since they are intended to give an upper limit, realistic 
estimate and lower limit, respectively,  as to how much 
of a problem the lensing effect will turn out to be
for future CMB experiments.
Apart from a factor of $\ln 10$,
the magnification fluctuation on a given scale
is essentially the area under a plotted curve out to
that scale. For instance, the {\rms} 
magnification fluctuations after 
smoothing on a scale of 5 arcminutes is about $15\%$
for the pessimistic model, where the sources that have
the dominant effect on the CMB are
assumed to be at redshifts $\gg 1$.
However, what matters for our purposes is of course 
not this {\rms} magnification fluctuation, 
but $\Clmag$ itself. In the pessimistic scenario, 
the maximum power is  
$\Clmag\sim \tento{-6}$, attained for $\l\sim 10^2$,
which ensures that the {\rms} lensing fluctuations are 
below the standard Poisson fluctuations for any reasonable 
flux cut.

\subsection{The uncertain point source normalization}

This conclusion in independent of
the overall normalization of the 
point source counts $\nbar'(\flux)$, since 
\Eq{PowerEq3} shows that the {\it ratio} of our lensing 
effect to the shot noise power $\Cps$
is independent of this normalization.
However, to assess the overall importance of both effects
to future CMB experiments, 
accurate knowledge of the radio source counts at the
frequencies where the observations are made is crucial.
Unfortunately, the point source population between
20 and a few hundred GHz is still
very poorly determined (see {\eg} Gundersen {\etal} 1997),
leaving open only the unattractive option of 
extrapolating from lower frequencies.
In \fig{ClFig} we have extrapolated the 1.5 GHz 
VLA FIRST data to 30 GHz by assuming that
the temperature fluctuations 
$(\Cps)^{1/2}(\nu)\propto\nu^{-(2+\alpha)}$.
To reflect the observational uncertainty, we  
have allowed $\alpha$ to vary as shown in Table 1,
with the middle-of-the-road estimate being that of
TE96.

\section{CONCLUSIONS}

We have calculated the effect of weak gravitational lensing on 
the power spectrum of unresolved point sources. 
We found that this adds a new term,
which is the magnification power spectrum
times $\nbar'(\fluxc)^2\fluxc^4$.
The ratio between this term and the standard Poisson 
term is roughly $(\N/4\pi)\Clmag$, independent of
the (poorly known) point source normalization, 
so if we keep shrinking the point source region in 
\fig{EverythingFig} by lowering the flux cut 
{\it ad infinitum}, the lensing term will eventually become
the dominant problem.
The lensing term initially {\it increases} as we remove more sources, 
reaching a maximum when $\sim 4$ sources per square degree are removed.

\Fig{EverythingFig} shows that for high-resolution interferometric 
experiments operating at low frequencies, a fairly aggressive source 
removal scheme will be necessary, bringing the
lensing effect close to this maximum strength.
Effective point source removal will be important for the
future Satellite missions {\it MAP} and {\it Planck} as well.
Fortunately, this strength does not exceed 12\% of the
standard Poisson fluctuations even in the worst-case scenario.

In summary, although this effect should be taken into account
in the detailed analysis of some future CMB experiments, 
it is likely to be a relatively small correction (like, e.g., the
Rees-Sciama effect) and 
should not substantially reduce the accuracy with
which cosmological parameters
can be measured.

\bigskip
The authors wish to thank Avi Loeb for 
suggesting this project and for helpful comments on the 
manuscript.
This work was partially supported by European Union contract
CHRX-CT93-0120, by Deutsche Forschungsgemeinschaft grant SFB-375
and by
NASA through a Hubble Fellowship,
{\#}HF-01084.01-96A, awarded by the Space Telescope Science
Institute, which is operated by AURA, Inc. under NASA
contract NAS5-26555.



\section{REFERENCES}


\rn Bersanelli, M. {\etal} 1996, {\it COBRAS/SAMBA
report on the phase A study}, ESA report D/SCI(96)3.

\rf Bond, J. R. \& Efstathiou, G. 1984;ApJ;285;L45

\rn Bond, J. R., Efstathiou, G. \& Tegmark, M. 1997, preprint astro-ph/9702100

\rf Broadhurst, T. J., Taylor, A. N. \& Peacock, J. A. 1995;ApJ;438;49

\rf Feldman, H. A., Kaiser, N. \& Peacock, J. A. 1994;ApJ;426;23

\rn Gundersen, J. O., Staveley-Smith, P., 
Payne, P. \& Lubin, P. M. 1997, submitted to {\it ApJL}.

\rf Jungman, G.. Kamionkowski, M., Kosowsky, A \& 
Spergel, D. N. 1996;Phys. Rev. D;54;1332

\rf Seljak, U. 1996;ApJ;463;1

\rg Tegmark, M. \& Efstathiou, G. 1996;MNRAS;281;1297;TE96

\rf Turner, E., Ostriker, J. P. \& Gott, J. R. III 1984;ApJ;284;1

\rf Villumsen, J. V. 1996;MNRAS;281;369

\rf Windhorst, R. A., Miley, G. K., Owen, F. N., Kron, R. G. 
\& Koo, D. C. 1985;ApJ;289;494

\rf Windhorst, R.A., Fomalont, E.B., Partridge, R.B., \& 
Lowenthal 1993;ApJ;405;498

\rn White, R. L., Becker, R. H., Helfand, D. J. \& 
Gregg, M. D. 1996, submitted to {\it ApJ}.

\rn Zaldarriaga, M., Spergel, D. \& Seljak, U. 1997, 
preprint astro-ph/9702157
       
 
\def\fheight{10.3cm} \def\fwidth{14.5cm}

\clearpage

\clearpage

\clearpage

\end{document}